\documentclass{naturemodified}
\usepackage{amsmath,amssymb,mathrsfs,latexsym,graphicx,psfrag}

\newcommand{\bs}[1]{\boldsymbol{#1}}

\newcommand{\ket}[1]{\left|#1\right\rangle}

\renewcommand{\eqref}[1]{Eq.~(\ref{#1})}

\def\ie{\emph{i.e.},\ }

\def\ea{\emph{et al.}}

\title{Statistical Phases and Momentum Spacings for One-Dimensional Anyons}
\author{Martin Greiter$^{*}$}

\begin{document}
\maketitle

\begin{affiliations}
 \item[] Institut f\"ur Theorie der
  Kondensierten Materie, Universit\"at Karlsruhe, D 76128 Karlsruhe
 \item[*] email: greiter@tkm.uni-karlsruhe.de
\end{affiliations}

\begin{abstract}
  Anyons and fractional statistics\cite{wilczek90,khare05} are by now
  well established in two-dimensional systems.  In one dimension,
  fractional statistics has been established so far only through
  Haldane's fractional exclusion principle\cite{haldane91prl937}, but
  not via a fractional phase the wave function acquires as particles
  are interchanged.  At first sight, the topology of the configuration
  space appears to preclude such phases in one dimension.  Here we
  argue that the crossings of one-dimensional anyons are always
  unidirectional, which makes it possible to assign phases
  consistently and hence to introduce a statistical parameter
  $\bs{\theta}$.  The fractional statistics then manifests itself in
  fractional spacings of the single-particle momenta of the anyons
  when periodic boundary conditions are imposed.  These spacings are
  given by $\bs{\Delta p =2\pi\hbar/L\,
    (|\theta|/\pi+\text{non-negative}}$ $\bs{\text{integer})}$ for a
  system of length $\bs{L}$.  This condition is the analogue of the
  quantisation of relative angular momenta according to
  $\bs{l_\text{z}=\hbar(-\theta/\pi+2\cdot\text{integer})}$ for
  two-dimensional anyons.
\end{abstract}

The concept of fractional statistics, 
as introduced by Leinaas and Myrheim\cite{leinaas-77ncb1}
and Wilczek\cite{wilczek82prl957}, has
generically been associated with identical particles in \emph{two
  space dimensions}.  It is intimately related to the topology of the
configuration space, or the existence of fractional relative angular
momentum.  Angular momentum does not exist in one dimension (1D), and
is quantised in units of $\hbar/2$ in three dimensions, due to the
commutation relations of the three generators of rotations.  In two
dimensions (2D), however, there is only one generator, $L_\text{z}$,
which may have arbitrary eigenvalues $l_\text{z}$.  Wilczek proposed
that two-dimensional anyons with statistical parameter $\theta$ and
relative angular momenta
$l_\text{z}=\hbar(-\theta/\pi+2\cdot\text{integer})$ may be realized
by particle flux-tube composites, attaching magnetic flux
$\Phi=2\theta\hbar c/e=\theta/\pi\cdot\Phi_0$ to bosons of charge $e$.
The choices $\theta=0$ and $\theta=\pi$ correspond to bosons and
fermions, respectively.

More fundamentally, the possibility of fractional statistics arises in
2D because one can associate a winding number with paths interchanging
particles.  The sum over paths in the many-particle path integral
consists of infinitely many topologically distinct sectors, which
correspond to the different winding configurations of the particles
around each other.  By the rules of quantum mechanics, one is allowed
to assign different weights to distinct sectors, provided these
weights satisfy the composition principle.  In particular, one may
assign a phase factor $e^{\pm i\theta}$ for each (counter-) clockwise
interchange of two particles.  This choice corresponds to Abelian
anyons with statistical parameter $\theta$ if the bare particles are
bosons.  The implicit assumption that the world lines never cross, \ie
the particles do not pass through each other, holds automatically for
all values $\theta\neq 0\ \text{mod}\ 2\pi$ due to the non-vanishing
relative angular momentum alluded to above.  In three or higher
dimensions, the only topologically inequivalent sectors correspond to
interchanges of particles, and the only consistent choices for the
statistics are bosons and fermions.  In 1D, the situation is alike if
particles are allowed to pass through each other, and trivial if they
are not.  In either case, \emph{the topology appears to preclude the
  possibility of one-dimensional anyons.}


The association of anyons with 2D, however, was challenged by
Haldane\cite{haldane91prl937} in 1991, who generalised the notion of
fractional statistics to arbitrary dimensions by defining statistics
through a fractional and hence generalised Pauli exclusion principle.
According to his definition, the statistics of anyons is given by a
rational ``exclusion'' parameter $g=p/q$ (with $p,q$ integer) which
states that the creation of $q$ anyons reduces the number of single
particle states additional anyons could be placed in by $p$.  In
particular, Haldane showed that the creation of $m$ quasiholes in a
$\nu=1/m$ Laughlin state\cite{laughlin83prl1395} reduces the number
of available single-quasiparticle states by $1$, which implies
$g=1/m$.  This result is fully consistent with the statistical
parameter $\theta=\pi/m$ obtained by Halperin\cite{halperin84prl1583}
and Arovas \ea\cite{arovas-84prl722}.

Most strikingly, however, Haldane showed that the spinons in the
Haldane--Shastry model (HSM)\cite{haldane88prl635,
shastry88prl639,haldane91prl1529,haldane-92prl2021}%
%
, a spin 1/2 chain with Heisenberg interactions which fall off as 
$1/r^2$ with the distance, 
obey half-Fermi exclusion statistics.  Haldane observed that for a
chain with $N$ sites, the number of single-particle states available
to additional spinons is given by $M+1$, where $M$ is the number of up
or down spins in the uniform singlet liquid, which in the presence of
$N_{\text{sp}}$ spinons is given by $M=(N-N_{\text{sp}})/2$.  The
creation of 2 spinons hence reduces the number of available states by
1, which implies $g=1/2$.  
%
(Note that since there are always fewer single-spinon states as there
are sites, localised spinon states cannot be orthogonal.)
Haldane further demonstrated that the dimension of the Hilbert space
spanned by the ground state and all possible many-spinon eigenstates
of the model is $2^N$, as required for a spin 1/2 system with $N$
sites.  The concept of fractional statistics hence was established in
a one-dimensional system, but it appeared that it could only be
defined through an exclusion principle.
%
Moreover, Haldane\cite{haldane91prl937} observed that the two
definitions of statistics do not always match, as hard-core bosons in
2D with magnetic flux-tubes attached
would be classified as anyons according to winding phases, but as
fermions according to his exclusion principle.

Let us briefly summarise:  Fractional statistics is fundamentally
associated with phases the many body wave functions acquire as
particles are interchanged or wind around each other, and can hence,
by the rules of quantum mechanics for identical particles as we know
them, only exist in 2D.  Nonetheless, according to an
alternative definition in terms of a generalised exclusion principle, 
fractional statistics can be defined independently of the dimension.
This alternative definition does not always match the original
one.  There would not be much reason to pay attention to it,
or even use the fractional exclusion of states as a definition of
fractional statistics, if there were not a concrete example of a
one-dimensional system (the HSM) which supports excitations with,
at least according to this definition, fractional statistics.

\begin{figure}
  \begin{center}
    \begin{minipage}[c]{0.7\columnwidth}
      \includegraphics[width=\linewidth]{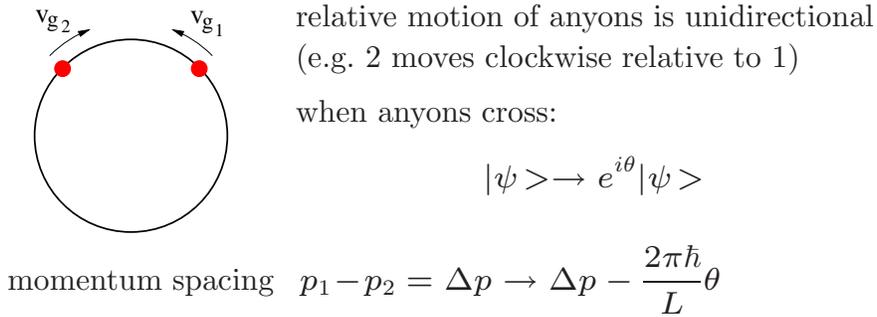}
    \end{minipage}
  \end{center}
  \vspace{\baselineskip}
  \caption{{\bfseries Fractional statistics in two and in one
      dimension.}  In 2D, a fractional phase $\theta$ acquired when
    anyons are interchanged through winding around each other
    manifests itself in a fractional shift in the relative angular
    momentum.  In 1D, a fractional phase when anyons cross manifests
    itself in a fractional shift in momentum spacing.  Consistency
    requires that the relative motion of 1D anyons is unidirectional,
    \ie that they always cross in the same direction.}
\end{figure}

In this Letter, we resolve the apparent conflict between the two
definitions.  The argument consists of several parts.  First, we show
that in the one-dimensional system obeying a fractional exclusion
principle, the HSM, an analog of a winding phase, \ie \emph{a
  statistical phase acquired by the wave function as the anyons go
  through each other}, exists.  The conflict with the topological
considerations explained above is circumvented in that the crossing of
the spinons occurs in one direction only.  The statistical phase of
$\pi/2$ acquired by the wave function as the spinons cross manifests
itself in a \emph{fractional shift for the spacings of the
  single-spinon momenta}.

Second, we show that a fractional shift for the momentum spacings, and
hence a statistical phase of $\pi/2$ acquired by the wave function,
also exists for the holons in the Kuramoto--Yokoyama model
(KYM)\cite{kuramoto-91prl1338}, the supersymmetrically extended HSM
allowing for itinerant holes.  
This suggests that the holons are half-fermions, a conclusion reached
previously by Ha and Haldane\cite{ha-94prl2887} using the asymptotic
Bethe ansatz (ABA), by Kuramoto and Kato\cite{kuramoto-95jpsj4518}
from thermodynamics, and by Arikawa, Saiga, and
Kuramoto\cite{arikawa-01prl3096} from the electron addition spectral
function of the model.
Since the $N$ localised single-holon states of the KYM are orthogonal,
however, they appear to be fermions according to Haldane's exclusion
statistics.  As a resolution of the conflict, we propose that the
exclusion principle yields the correct statistics only when applied to
energy eigenstates of a given model.
%

Finally, we argue that the picture we propose---crossings in
only one direction, statistical phases acquired by the wave function as
anyons go through each, fractionally spaced single anyon
momenta---holds for 1D anyons in general.

The subtleties involved are best explained by looking closely at
two-spinon and two-holon eigenstates of the KYM.  The model is
conveniently formulated by embedding the one-dimensional chain with
PBCs into the complex plane by mapping it onto the unit circle with
the sites located at complex positions
$\eta_\alpha=\exp\!\left(i\frac{2\pi}{N}\alpha\right)$, where $N$
is the number of sites and $\alpha=1,\ldots,N$.  Each site can be
occupied either by an up- or down-spin electron or a hole (\ie the
site is empty).  The Hamiltonian is given by
\begin{equation}
  H_{\text{KY}}=-\frac{2\pi^2}{N^2}
  \sum_{\alpha\neq\beta}^{N}
  \frac{P_{\alpha \beta}}{\vert \eta_{\alpha} - \eta_{\beta} \vert^2},
  \label{kyham}
\end{equation}
where the graded permutation operator $P_{\alpha \beta}$ exchanges
particles on sites $\eta_\alpha$ and $\eta_\beta$, multiplied by a
minus sign if both particles are fermions (\ie neither of them a hole).
In the absence of holes, \eqref{kyham} reduces to the HSM, which
possesses the exact ground state
\begin{equation}
  \label{gswave}
  \Psi_0[z_i] 
  =\prod_{i < j}^M (z_i-z_j)^2 \prod_{i=1}^M z_i
\end{equation}
for $N$ even, $M={N}/{2}$, and $[z_i]\equiv(z_1,\ldots,z_M)$.
The $z_i$'s denote the positions of the up spins.  
%
The greatly simplifying feature of the HSM (and the
KYM) is that the spinons (and the holons) are free in the sense that
they only ``interact'' through their half-Fermi
statistics\cite{ha-93prb12459,essler95prb13357,greiter-05prb224424}. 

Let us now turn to the two-spinon eigenstates.  A momentum basis for
spin-polarised two-spinon states is given by
\begin{equation}
  \label{psinm}
  \Psi_{mn}[z_i] 
  =\sum_{\alpha,\beta}^N (\bar\eta_\alpha)^m (\bar\eta_\beta)^n
  \prod_{i=1}^M(\eta_{\alpha}-z_{i})(\eta_{\beta}-z_{i})\,\Psi_0[z_i],
\end{equation}
where $M=(N-2)/2$ and $M\ge m\ge n\ge 0$.  For $m$ or $n$ outside this
range, $\Psi_{mn}$ vanishes identically, reflecting the
overcompleteness of the position space basis.  Acting with
\eqref{kyham} on \eqref{psinm} yields\cite{bernevig-01prl3392}
\begin{equation}
  \label{scattspinon}
  H_{\text{KY}}\ket{\Psi_{mn}}=
  E_{mn}\ket{\Psi_{mn}}+\sum_{l=1}^{l_{\text{max}}}V_l^{mn}\ket{\Psi_{m+l,n-l}}
\end{equation}
with $l_{\text{max}}\!=\!\min(M\!-\!m,n)$,
$V_l^{mn}\!=\!-\frac{2\pi^2}{N^2}(m\!-\!n\!+\!2l)$,
and 
\begin{equation}
  \label{etotspinon}
  E_{mn}=E_0+\epsilon(q_m)+\epsilon(q_n).
\end{equation}
$E_0=-\frac{\pi^2}{4N}$ is the ground state energy,
\begin{equation}
  \label{epsilon}
  \epsilon(q)=\frac{1}{2}\,{q\,(\pi-q)}+\frac{\pi^2}{8N^2},
\end{equation}
and we have identified the single-spinon momenta for $m\ge n$
according to
\begin{equation}
  \label{qmqnspinon}
  q_m\!=\!\pi\!-\!\frac{2\pi}{N}\!\left(m\!+\!\frac{1}{2}\!+\!s\right)\!,\
  q_n\!=\!\pi\!-\!\frac{2\pi}{N}\!\left(n\!+\!\frac{1}{2}\!-\!s\right)\!,
\end{equation}
with a \emph{statistical shift} $s=1/4$.
Since the ``scattering'' of the non-ortho\-gonal basis states
$\ket{\Psi_{mn}}$ in \eqref{scattspinon}  only occurs in one direction, 
increasing $m-n$ while keeping $m+n$ fixed, the eigenstates of 
$H_{\text{KY}}$ have energy eigenvalues $E_{mn}$. 

The relevant feature for our present purposes is the shift $s$ in the
single-spinon momenta \eqref{qmqnspinon}, which we will elaborate on
now.  The state \eqref{psinm} tells us unambiguously that the sum of
both spinon momenta is given by
$q_m+q_n=2\pi-\frac{2\pi}{N}(m+n+1)$, and hence \eqref{qmqnspinon}.
The shift $s$ is determined by demanding that the excitation energy
\eqref{etotspinon} of the two-spinon state is a sum of single-spinon
energies, which in turn is required for the explicit solution here to
be consistent with the ABA 
results\cite{ha-93prb12459,essler95prb13357,greiter-05prb224424}.

The appearance of this shift, which decreases the momentum $q_m$ of
spinon 1 and increases momentum $q_n$ of spinon 2, is somewhat
surprising, given that the basis states \eqref{psinm} are constructed
symmetrically with regard to interchanges of $m$ and $n$.  To
understand this asymmetry, note that $M\ge m\ge n\ge 0$ implies
$0<q_m<q_n<\pi$.  The dispersion \eqref{epsilon} implies that the
group velocity of the spinons is given by
\begin{equation}
  v_\text{g}(q)=\partial_q\epsilon(q)=\frac{\pi}{2}-q,
\end{equation}
which in turn implies that $v_\text{g}(q_m)>v_\text{g}(q_n)$.  The
physical significance of this result can hardly be overstated.  It
means that the \emph{relative motion} of spinon 1 (with $q_m$) with
respect to spinon 2 (with $q_n$) is \emph{always counterclockwise} on
the unit circle.  Then, however, the shifts in the individual spinon
momenta can be explained by simply assuming that the two-spinon state
acquires a statistical phase $\theta=2\pi s$ whenever the spinons pass
through each other.  This phase implies that $q_m$ is shifted by
$-\frac{2\pi}{N}s$ since we have to translate spinon 1
counterclockwise through spinon 2 and hence counterclockwise around
the unit circle when obtaining the allowed values for $q_m$ from the
PBCs.  Similarly, $q_n$ is shifted by $+\frac{2\pi}{N}s$ since we have
to translate spinon 2 clockwise through spinon 1 and hence clockwise
around the unit circle when obtaining the quantisation of $q_n$.
%
(The fact that the ``bare'' ($s=0$) values for $q_m$ and $q_n$ are
quantised as $\frac{2\pi}{N}\bigl(\frac{1}{2}+\text{integer}\bigr)$ is
related to the bosonic representation of the ``bare'' spinons.  If we
had chosen a fermionic representation, they would be quantised as
$\frac{2\pi}{N}\cdot\text{integer}$.)

That the crossing of the spinons occurs only in one direction is not
just a peculiarity, but a necessary requirement for fractional
statistics to exist in 1D at all.  If the spinons could
cross in both directions, the fact that paths interchanging them twice
(\ie once in each direction) are topologically equivalent to paths not
interchanging them at all would imply $2\theta=0\ \text{mod}\ 2\pi$
for the statistical phase, \ie only allow for the familiar choices of
bosons or fermions.  With the scattering occurring in only one
direction, arbitrary values for $\theta$ are possible.  The
one-dimensional anyons neither break time-reversal symmetry (T) nor
parity (P).

We now turn to the two-holon eigenstates of the
KYM\cite{thomale-06prb024423}, which are highly instructive
with regard to Haldane's exclusion principle as a definition of
fractional statistics.  A momentum basis for two-holon states is given
by
\begin{equation}
  \label{twoholewave}
  \Psi_{mn}^{\text{ho}}[z_i,h_j]
  =\phi_{mn}(h_1,h_2) 
  \prod_{i=1}^M (h_1-z_i)(h_2-z_i)\Psi_0[z_i], 
\end{equation}
where $M=(N-2)/2$ and $[z_i,h_j]\equiv(z_1,\ldots,z_M;h_1,h_2)$.  The
$z_i$'s denote the positions of the up spins again, and $h_1,h_2$ the
positions of the holes.  $\phi_{mn}(h_1,h_2)$ is an internal
holon-holon wave function.
Using the educated guess
$
  \phi_{mn}(h_1,h_2)=(h_1-h_2)(h_1^m h_2^n+h_1^n h_2^m), 
$
we obtain 
\begin{equation}
  \label{scattholon}
  H_{\text{KY}}^{\text{ho}}\ket{\Psi_{mn}^{\text{ho}}}=
  E_{mn}^{\text{ho}}\ket{\Psi_{mn}^{\text{ho}}}
  +\sum_{l=1}^{l_{\text{max}}}
  V_l^{mn}\ket{\Psi_{m-l,n+l}^{\text{ho}}}
\end{equation}
for $0 \le n \le m \le M+1$.  If this condition is violated,
the basis states $\ket{\Psi_{mn}^{\text{ho}}}$ do not vanish
identically, but it is not possible to construct eigenstates from
them.  In \eqref{scattholon}, $l_{\text{max}}$ is the largest
integer $l\le\frac{m-n}{2}$,
\mbox{$V_l^{mn}=\frac{2\pi^2}{N^2}(m-n)$,} and
\begin{equation}
  \label{etotholon}
  E_{mn}^{\text{ho}}=E_0+\epsilon^{\text{ho}}(p_m)+\epsilon^{\text{ho}}(p_n). 
\end{equation}
The single-holon energies are given by
\begin{equation}
  \label{epsilonholon}
  \epsilon^{\text{ho}}(p)=\frac{1}{2}\,{p\,(\pi+p)}-\frac{\pi^2}{8N^2},
\end{equation}
and we have identified the single-holon momenta for \mbox{$m\ge n$}
according to
\begin{equation}
  \label{qmqnholon}
  p_m=-\pi+\frac{2\pi}{N}\left(m\!+\!s\right),\
  p_n=-\pi+\frac{2\pi}{N}\left(n\!-\!s\right),
\end{equation}
with $s=1/4$.  The ``scattering'' occurs again only in one direction,
this time decreasing $m-n$ while keeping $m+n$ fixed, which implies
both that the basis states $\ket{\Psi_{mn}^{\text{ho}}}$ are not
orthogonal and that the two-holon eigenstates of $H_{\text{KY}}$ have
energy eigenvalues $E_{mn}^{\text{ho}}$.  The statistical shift $s$ is
once again determined by demanding that the holons are free, which in
turn is required by consistency with the ABA
solutions\cite{essler95prb13357}.

The momenta are again limited to about half of the Brillouin zone,
$-\pi\!-\!\frac{\pi}{2N}\le p_n < p_m \le\frac{\pi}{2N}$.  With the holon
group velocity
\begin{equation}
  v_\text{g}^{\text{ho}}(p)=\partial_p\epsilon^{\text{ho}}(p)=\frac{\pi}{2}+p,
\end{equation}
we obtain $v_\text{g}^{\text{ho}}(p_m) > v_\text{g}^{\text{ho}}(p_n)$.
The crossing of the holons occurs again only in one direction, and
the momentum shifts as well as the half-Fermi statistics emerges as in
the case of the spinons, except that the state now acquires
the phase $\theta=-2\pi s$, with the result that the momentum $p_m$ of
the holon with the larger group velocity $v_\text{g}^{\text{ho}}(p_m)$
is shifted by $+\frac{2\pi}{N}s$, and $p_n$ shifted by
$-\frac{2\pi}{N}s$.  Physically, this reversal in the sign reflects
that the holon is created by annihilation of an electron at a spinon
site, \ie by removing a fermion from a half-fermion.  The spacing
between $p_m$ and $p_n$, however, is quantised as for the spinons
above.
Note that the hard-core constraint of the holons is irrelevant here.

Let us now reconcile this result with the exclusion principle.  As
mentioned, the hard-core condition for holons effects that they are
fermions according to Haldane's exclusion principle applied to states
localised in position space.  When applied to exact eigenstates of the
model, however, the result is different.  Since the creation of 2
holons decreases the number of up or down spins in the uniform liquid
$M$ by 1, the number of single-holon states (labelled by $m$ or $n$
above) available for additional holons decreases by 1.  This implies
half-Fermi statistics, and is consistent with the momentum spacings.
%
\emph{The exclusion principle hence yields the correct statistics only if
applied to eigenstates of the model.}  
The wave function for localised holons
is really a superposition of a holon state (onto which we project in
\eqref{twoholewave}) and a holon surrounded by an incoherent spinon
cloud in a singlet configuration. 

So far, our discussion has been limited to a particular model.  The 
conclusions, however, hold in general.  As noted
above, the KYM is special in that the spinon and holon excitations
are free.  The single spinon and holon momenta are hence good quantum
numbers.  The eigenstates of the model can be labelled in terms of
these momenta, which we have shown to be fractionally spaced.  Any
other model of a one dimensional spin chain can be described as a
KYM supplemented by additional terms, which give rise to an interaction
between the spinons and holons.  This interaction will scatter the
basis states of free spinons and holons, the 
eigenstates of the KYM, into each other.  The eigenstates of the
interacting model will hence be superpositions of states with
different single spinon and holon momenta, all of which, however, will
be fractionally spaced.
In other words, the fractional shifts in \eqref{qmqnspinon},
\eqref{qmqnholon} (and also \eqref{angmom}, \eqref{linmom} below)
will still be good quantum numbers, while the integers $n$ and $m$
will turn into ``superpositions of integers''.

This argument shows that whenever we have spinons and holons in a
one-dimensional spin chain, we have fractionally spaced single
particle momenta as a consequence of their fractional statistics.  Is
it reasonable to assume that this picture holds for anyons in 1D in
general?  We believe there are very good reasons to do so.  First,
spinons and holons are the only known examples of anyons in 1D.  This
picture hence holds for all examples of 1D systems with fractional
statistics.  Second, the picture resolves a profound conflict, as
topology precludes the existence of one dimensional anyons in a
conventional framework of indistinguishable particles.  The conflict
is circumvented here in that the anyons become distinguishable through
their dynamics, and cross in one direction only.  If the picture we
propose here were not of general validity, another resolution to this
conflict would have to exist.  This does not appear to be the case.
In any event, the picture we propose is the only consistent picture
available at present.  It is hence only reasonable to assume general
validity.

We conclude with a summary.  We propose that the statistics of
identical particles is always reflected in the quantisation condition
of an observable quantity.  For anyons with statistical parameter
$\theta$ in 2D, the kinematical relative angular momentum
between two anyons is quantised as\cite{wilczek82prl957}
\begin{equation}
  \label{angmom}
  l_\text{z}=\hbar\left(-\frac{\theta}{\pi}+2m\right),
\end{equation}
where $-\pi<\theta\le\pi$ and $m$ is integer.

For anyons with statistical parameter $\theta$ in a one-dimensional
system with length $L$ and periodic boundary conditions---and this is
the central message of this Letter---the allowed values for the
spacings between the kinematical (linear) momenta are quantised as
\begin{equation}
  \label{linmom}
  p_{i+1}-p_i=\Delta p=\frac{2\pi\hbar}{L}\left(\frac{|\theta |}{\pi}+n\right)
\end{equation}
for $p_{i+1}-p_i\ge 0$, where $-\pi<\theta\le\pi$ and $n$ is a
non-negative integer.  The spacing condition \eqref{linmom} holds for
many-anyon states with single-particle momenta $p_1\le p_2\le\ldots\le p_N$
in any interval $p_i\in\mathcal{I}$, provided that the anyon group
velocity $v_\text{g}(p)=\partial_p\epsilon(p)$ is a strictly 
increasing ($\theta<0$) or decreasing ($\theta>0$) function of $p$ in
this interval.  This condition is required for the anyons to cross
in one direction only.
%
In an interacting many particle system, the quantum numbers 
$m$ and $n$ in \eqref{angmom} and \eqref{linmom} are not expected to be
good quantum numbers.  The fractional shifts $-{\theta}/{\pi}$ and
${|\theta |}/{\pi}$, however, are topological invariants.

Note that \eqref{angmom} and \eqref{linmom} hold only between the
\emph{physical} or \emph{kinematical} statistics of the anyons and the
\emph{kinematical} angular or linear momenta, as canonical momenta are
gauge dependent.  In particular, one may change the canonical momenta
while simultaneously changing the canonical statistics of the fields
(\ie the statistics imposed when canonically quantising the fields)
used to describe the anyons via a ``singular'' gauge transformation.
%
The canonical statistics may either be bosonic, as in the case of the
spinons in the analysis above, or fermionic, as in the case of the
holons above.

%
Our analysis further demonstrates that particular care must be
exercised when defining statistics using Haldane's exclusion
principle.  The fact that it gives the correct result for the
statistics of holons in the KYM when applied to eigenstates of the
model but an incorrect result when applied to holon states localised
in position space leads us to conjecture that in general,
\emph{the exclusion principle yields correct results only when applied
 to eigenstates of a given model}. 


\begin{addendum}
\item I wish to thank D.~Schuricht and R.~Thomale for helpful
  suggestions on the manuscript.
\item[Competing Interests] The author declares that he has no
  competing financial interests.
\end{addendum}

\end{document}